\documentclass[10pt,twocolumn,twoside]{IEEEtran}
\usepackage{amsmath}
\usepackage{amssymb}
\usepackage{algorithmicx}
\usepackage[utf8]{inputenc}  
\usepackage{textcomp}        
\usepackage{algpseudocode}
\usepackage{graphicx}
\usepackage{subfigure}
\usepackage{epstopdf}
\usepackage{cite}
\usepackage{textcomp}
\usepackage{mathrsfs}
\usepackage{color}
\usepackage{booktabs}
\usepackage{cases}
\usepackage{setspace}
\usepackage{bm}
\usepackage{cuted}
\usepackage{booktabs}
\usepackage{stfloats}
\usepackage{makecell}
\usepackage{mathtools}
\usepackage[ruled,linesnumbered]{algorithm2e}

\begin{document}
\title{\Huge Movable-Antenna Empowered Backscatter ISAC: Toward Geometry-Adaptive, Low-Power Networks}
	\author{ 
				Haohao~Zhang, Bowen~Gu,  Xianhua~Yu, Hao~Xie, Liejun~Wang, Yongjun~Xu, \textit{Senior Memeber, IEEE},\\              
                Xiaoming Tao, \textit{Senior Memeber, IEEE},  Haijun Zhang, \textit{Fellow, IEEE}
				\IEEEcompsocitemizethanks{			
				\IEEEcompsocthanksitem H. Zhang, B. Gu, and L. Wang are with the School of Computer Science and Technology, Xinjiang University, Urumqi, Xinjiang 830049, China, and with Xinjiang Multimodal Intelligent Processing and Information Security Engineering Technology Research Center, Urumqi, Xinjiang 830049, China (e-mails: hhzhang@stu.xju.edu.cn, bwgu@xju.edu.cn, wljxju@xju.edu.cn).
                \IEEEcompsocthanksitem X. Yu and H. Xie are with the School of Electrical Engineering and Intelligentization, Dongguan University of Technology, Dongguan, China (e-mails: xianhuacn@foxmail.com, xiehao@dgut.edu.cn).
                \IEEEcompsocthanksitem Y. Xu is with the School of Communications and Information Engineering, Chongqing University of Posts and Telecommunications, Chongqing 400065, China (e-mail: xuyj@cqupt.edu.cn).
                \IEEEcompsocthanksitem X. Tao  is with the School of Computer Science and Technology, Xinjiang University, Urumqi, Xinjiang 830049, and also with the Department of Electronic Engineering, Tsinghua University, Beijing, 100084, China  (e-mail: taoxm@tsinghua.edu.cn)
                \IEEEcompsocthanksitem H. Zhang is with the School of Computer and Communication Engineering, University of Science and Technology Beijing, Beijing 100083, China (e-mail: haijunzhang@ieee.org).
                    } \vspace{-10pt}	
	
	}


\maketitle

\begin{abstract}
Backscatter-based integrated sensing and communication (B-ISAC) elevates passive tags into information-bearing scatterers, offering an ultra-low-power path toward dual-function wireless systems. However, this promise is fundamentally undermined by a cascaded backscattering link that suffers from severe double fading and is exquisitely sensitive to geometric misalignment. This article tackles this geometric bottleneck by integrating movable antenna systems (MAS) at the transceiver side. MAS provides real-time, controllable spatial degrees of freedom through sub-wavelength antenna repositioning, enabling active reconfiguration of the cascaded channel without modifying passive tags or consuming additional spectrum. We position this solution within a unified ISAC–backscatter communication–B-ISAC evolution, describe the resulting MAS-assisted B-ISAC architecture and operating principles, and demonstrate its system-level gains through comparative analysis and numerical results. Finally, we showcase the potential of this geometry-adaptive paradigm across key IoT application scenarios, pointing toward future motion-aware wireless networks.
\end{abstract}


\section{Introduction}

The rapid advancement of wireless communications and electronic manufacturing is pushing the Internet of Things (IoT) from simple connectivity toward the intelligent interconnection of everything. In this emerging paradigm, billions of nodes are expected not only to exchange data, but also to sense, interpret, and interact with their surroundings. Applications such as autonomous driving, smart logistics, and industrial automation epitomize this trend: they require devices to support high-rate, ultra-reliable links and achieve centimeter-level perception and real-time situational awareness \cite{10608156}.

Integrated sensing and communication (ISAC) has emerged as a promising enabler of this dual capability by allowing a single waveform and hardware platform to jointly perform data transmission and sensing \cite{gonzalez2025six}. However, current ISAC realizations still rely on actively transmitted RF illumination, which introduces significant power consumption, synchronization overhead, and hardware complexity, making it difficult to sustain in dense, battery-free IoT deployments. Their coherent operation is also sensitive to rapidly varying multipath, making purely active ISAC architectures fragile in highly dynamic environments. These limitations motivate moving toward architectures that can reuse existing radio signals rather than continuously generating new sensing waveforms.

Backscatter-based ISAC (B-ISAC) represents an important step in this direction. Leveraging backscatter communication (BackCom), passive tags modulate and reflect an illuminating ISAC waveform, embedding their own data while implicitly encoding geometric cues of the tagged object \cite{zargari2023sensing}. Specially, a cooperative transmitter (e.g., a base station (BS)) transmits the unified ISAC signal, and tags attached to vehicles, goods, or wearables act as ultra-low-power information-bearing scatterers. The BS (or a companion receiver) then interprets the backscattered echo as a dual-purpose signal, simultaneously recovering the tag data and estimating range, velocity, or orientation. This shifts the RF burden to the infrastructure side while keeping tags battery-free and RF-chain-free, yielding a power profile naturally suited to large-scale, long-lived IoT deployments.  However, because both communication and sensing share the same forward-backward propagation path, the cascaded channel suffers from pronounced double-path attenuation, which severely limits received power and sensing range \cite{gu2024breaking}. The backscattered return is also highly geometry-dependent: small changes in tag orientation, user posture, or environmental blockage can disrupt either hop and trigger deep fading. Multi-antenna and intelligent reflecting surface (IRS) techniques can electronically adjust beam patterns, but they still operate over fixed physical apertures at fixed locations \cite{Xutcom}. They can redistribute energy within a given geometry, yet cannot alter the underlying spatial relationship among the BS, tag, and receiver. When dominant scatterers or tagged objects move or become blocked, purely electronic beam steering may still leave geometry-induced dead zones, highlighting the need for complementary transceiver-side mechanisms that can influence not only waveforms or beams, but the effective link geometry itself.

Movable antenna systems (MAS) provide a concrete way to realize such transceiver-side geometric control. The core question is simple yet fundamental: if antenna elements are allowed to reposition within a sub-wavelength region during the channel coherence time, can the geometry of the cascaded ISAC link be actively reshaped? By enabling fine-grained motion of transmit and/or receive antennas, MAS creates additional spatial degrees of freedom (DoFs) that allow the system to steer multipath illumination, escape deep fading, and follow dominant echo paths without modifying passive tags or consuming extra spectrum \cite{zhu2024movable}.  Transmit-side antenna motion helps align the incident field with the tag’s reflective state, whereas receive-side antenna motion optimizes the aperture’s position to capture the most informative returns. These motion-induced DoFs substantially strengthen backscatter detection and enhance sensing precision, turning link geometry from a fixed constraint into an adjustable design variable.

Building on this capability, MAS-empowered B-ISAC establishes a geometry-aware operating paradigm in which dynamic illumination and adaptive reception are jointly leveraged to support robust ISAC. 
Coordinated antenna motion enables the transceiver not only to observe the cascaded backscatter channel but also to subtly influence its effective structure, thereby exploiting the dual nature of the backscattered waveform as a carrier of both digital information and spatial cues \cite{fang2025integrated}. 
This leads to systems capable of maintaining reliable links and high-fidelity sensing under mobility, blockage, and complex environmental dynamics, which are difficult to achieve with static architectures. 
On this basis, this article presents the architecture, operating principles, and system-level capabilities of MAS-empowered B-ISAC and discusses its application prospects in future low-power IoT networks.
The main contributions are summarized as follows.

\begin{itemize}
    \item We first offer an integrated overview of ISAC, BackCom, and B-ISAC, highlighting how backscatter supports ISAC and identifying the structural limitations of B-ISAC.
    \item We then introduce a MAS-empowered B-ISAC architecture, explain how antenna motion reshapes the cascaded channel and joint sensing-communication behavior, and validate its advantages through comparative evaluation.
    \item Finally, we discuss representative application scenarios and outline key challenges and opportunities for practical, scalable deployment of MAS-assisted B-ISAC in future IoT environments.
\end{itemize}

\section{Evolution from ISAC to B-ISAC}

This section reviews the conceptual foundations and technical evolution from ISAC to BackCom, and finally to B-ISAC. We first revisit the design principles of ISAC and BackCom, then highlight how B-ISAC merges these paradigms and why its inherent geometric sensitivity motivates transceiver-side spatial adaptability.

\subsection{ISAC}

ISAC builds on the observation that a wireless waveform naturally interacts with its environment. Instead of separating communication and sensing, ISAC exploits this interaction so a single waveform can simultaneously convey data and extract spatial information \cite{10608156}. A typical ISAC system integrates three components: (i) a shared transceiver front-end that maintains mutual coherence across functionalities; (ii) dual-purpose waveforms that embed communication symbols while preserving radar-friendly delay-, Doppler-, and angle-sensitive structures; and (iii) joint signal processing pipelines that decode data and estimate spatial parameters from the same received signal \cite{luo2025isac}.  By jointly designing transceivers, waveforms, and processing pipelines, ISAC achieves high spectral and energy efficiency, allowing wireless systems to “see” and “talk” simultaneously.

This integration reduces hardware redundancy, improves spectral and energy efficiency, and simplifies deployment in large-scale intelligent environments. However, ISAC inherits limitations tied to active transmission: continuous illumination demands power amplifier activity, high-rate sampling, and precise synchronization, which hinder scalability to massive or battery-free IoT. Moreover, waveform requirements for communication and sensing can conflict, and maintaining coherent processing becomes difficult in dynamic, multipath-rich environments. These constraints reveal a tension between functionality and energy sustainability, motivating architectures that reuse existing radio signals and minimize active emissions.

\subsection{BackCom}

\begin{figure} [t]
    \centering
    \includegraphics[width=0.9\linewidth]{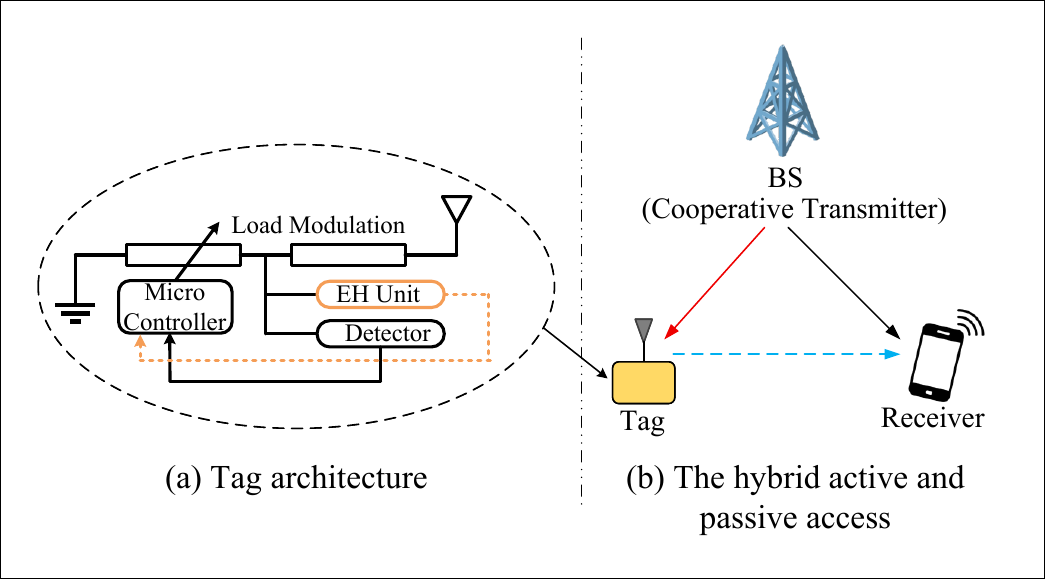}
    \caption{The main architectures of the hybrid active-passive system.}
    \label{fig1}
\end{figure}

BackCom enables a passive device, or small tag, to convey information by modulating the reflection of incident radio signals, avoiding the need for an active RF chain, making it an inherently ultra-low-power communication technique \cite{10129192}. As shown in Fig.~\ref{fig1}(a), a tag splits the received power into two functional paths: i) a load-modulation circuit, controlled by a microcontroller, toggles among discrete impedance states to produce reflection coefficients for encoding binary or higher-order symbols; ii) an energy-harvesting rectifier converts part of the incident power into DC to sustain the microcontroller, enabling battery-free and long-lived operation.

Traditional BackCom systems (monostatic or bistatic) rely on a dedicated carrier emitter colocated with or separate from the reader.  Ambient BackCom (AmBC) generalizes this mechanism by letting the tag modulate naturally occurring RF signals (e.g., Wi-Fi and Bluetooth) without any dedicated illuminator.  
AmBC significantly improves deployability and energy efficiency but faces additional technical challenges: i) fluctuating and unpredictable incident power, and ii) interference from multiple ambient RF sources, which complicate demodulation.

To mitigate these uncertainties while preserving the ultra-low-power nature of BackCom, recent systems adopt a hybrid active-passive (symbiotic) architecture, as illustrated in Fig.~\ref{fig1}(b), where a cooperative transmitter (e.g., a BS or a suitable access point (AP)) emits a controlled illumination waveform that simultaneously serves conventional communication and provides a carrier for passive tags \cite{gu2024breaking}. Since this  transmitter is already powered on for its primary downlink, reusing this waveform as an illuminator incurs only marginal extra energy, while tags remain battery-free and free of RF power amplifiers. By backscattering the same signal under shared spectrum and timing references, tags enable coherent decoding and naturally integrate into ISAC, with the transmitter waveform supplying delay/Doppler structure for sensing and the tags acting as controllable scatterers, thereby paving the way for B-ISAC without compromising device-side ultra-low-power operation.

\subsection{B-ISAC}

B-ISAC represents a targeted evolution of the ISAC paradigm by focusing on scenarios where the sensing targets are explicitly equipped with backscatter tags. Unlike classical ISAC, which interacts solely with passive and unstructured environmental reflections, B-ISAC introduces ultra-low-power tags that deliberately shape the returned waveform through load modulation \cite{tian2024performance}. Tagged objects such as vehicles, packages, or wearables thus become controllable, information-bearing scatterers, enabling the BS/AP to extract both digital data and geometric sensing features from the same reflected signal \cite{icc2024}.

\begin{figure}
    \centering
    \includegraphics[width=0.9\linewidth]{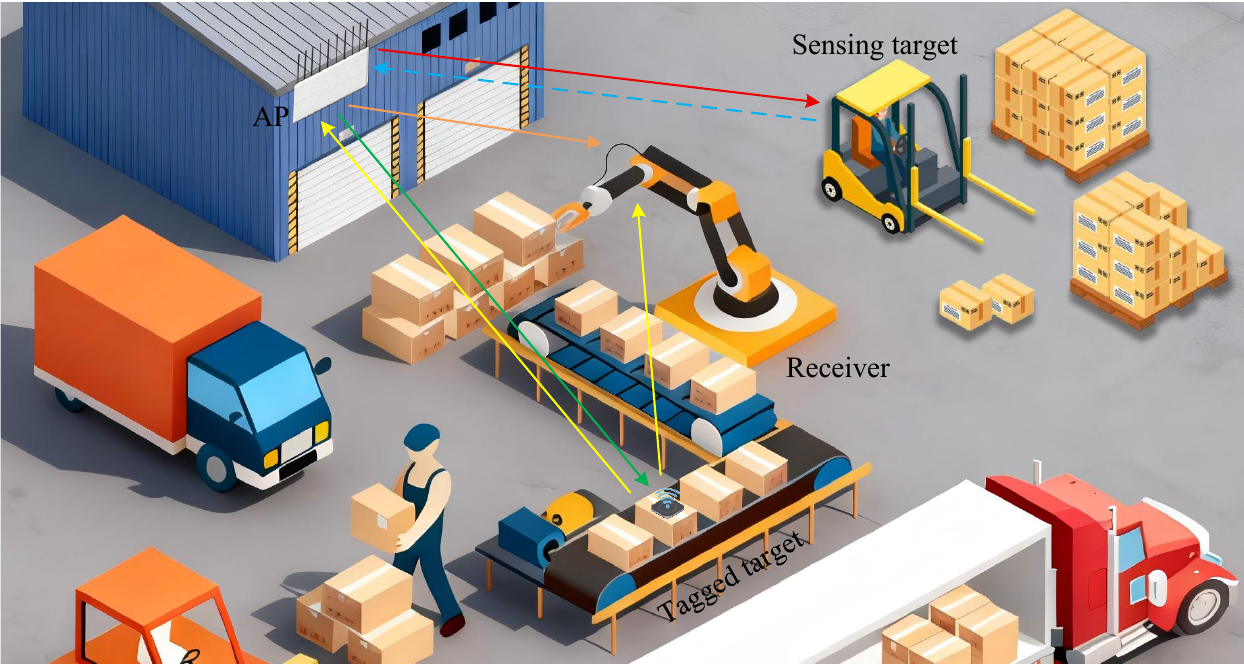}
    \caption{The architectural diagrams of traditional ISAC and B-ISAC.}
    \label{fig2}
\end{figure}

As illustrated in Fig.~\ref{fig2}, the full-duplex AP illuminates the tagged object with a unified ISAC waveform, while the tag performs a dual function: it embeds its symbols via impedance modulation and simultaneously returns a reflection whose delay, Doppler, and angular structure reveal the object’s position and motion \cite{yang2023novel}. Meanwhile, the active link from the BS to the receiver provides high-power reference signals and stable timing/synchronization, ensuring coherent processing of the weak backscatter component. In this way, hybrid active-passive operation enables B-ISAC to couple active illumination with passive modulation, enriching conventional ISAC with an additional layer of distributed, tag-enabled intelligence without compromising the ultra-low-power design of the tag side.

Despite these benefits, the physics of the cascaded forward-backward channel imposes intrinsic limitations.
The composite link suffers from double-path fading, making communication and sensing extremely sensitive to geometric alignment.
Unlike active ISAC, which can extract cues from multipath diversity, B-ISAC relies heavily on a sufficiently strong forward and backward lines-of-sight (LoS) component; even small orientation changes, blockages, or shadowing can push the link into deep fading.
Moreover, the tag’s reflective strength depends on its impedance state, orientation, and available harvested energy, causing the backscattered amplitude and phase to fluctuate over time.
Because the tag lacks active gain or beamforming capability, these variations cannot be compensated locally, and static transceivers or IRS arrays have limited ability to restore link robustness in dynamic environments.

These structural constraints imply that while B-ISAC is energy- and spectrum-efficient, it suffers from limited spatial controllability and geometric robustness. Its performance is fundamentally tied to fixed transceiver geometry, making it difficult to sustain reliable communication and sensing when the environment, channel, or target position changes rapidly. This, in turn, calls for new transceiver-side mechanisms that can actively adjust the transmitter-tag-receiver geometry itself, rather than passively enduring it.

\section{The MAS-empowered B-ISAC System}

To address the above challenges, MAS offers a compelling solution. By allowing the transmitter or receiver to physically reposition their antenna elements within a small region during the channel coherence time, MAS turns the propagation geometry from a static constraint into an active design dimension. This section introduces MAS fundamentals, their hardware realizations, and how MAS fundamentally reshapes B-ISAC operation and performance.

\subsection{MAS}

\begin{figure*} [t]
     \vspace{-5mm}
    \centering
    \includegraphics[width=0.65\linewidth]{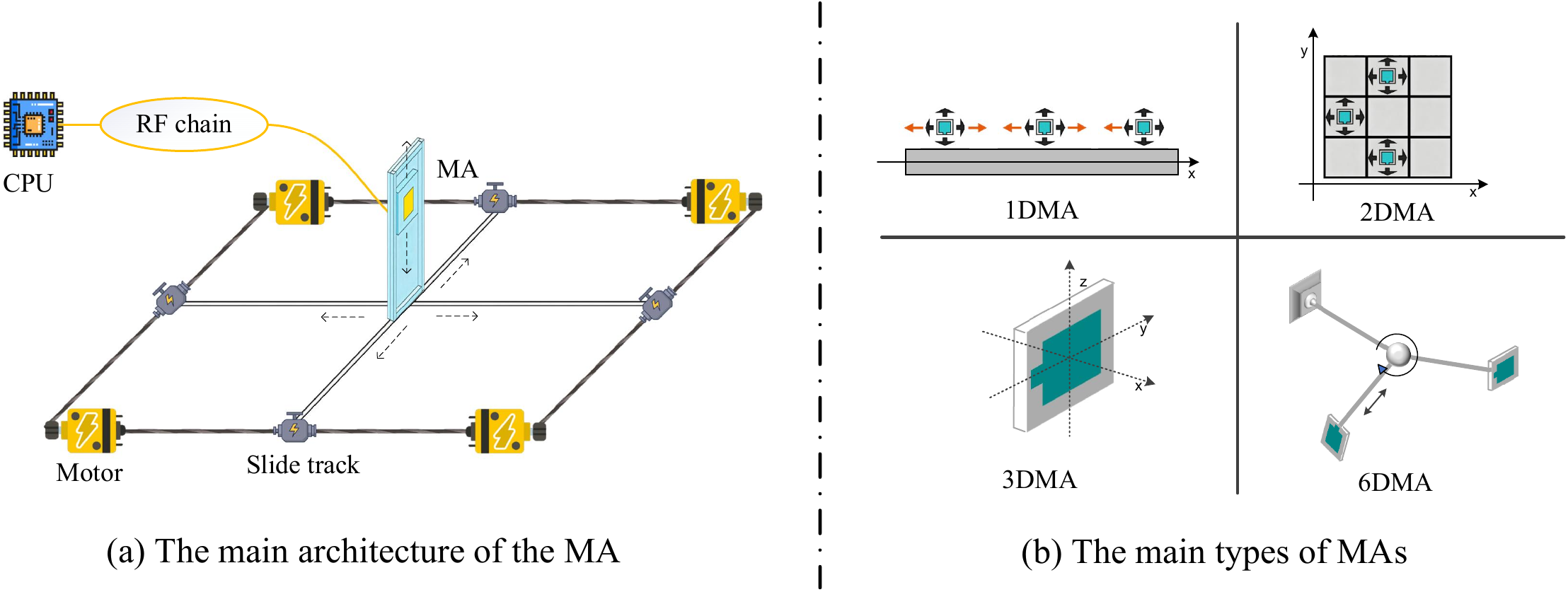}
    \caption{The main architecture and type of MAs.}
    \label{fig3}
\end{figure*}

Conventional antenna arrays rely on fixed apertures and fixed viewpoints. Once deployed, neither their effective aperture nor their interaction with the environment can change. MAS breaks this assumption. As illustrated in Fig.~\ref{fig3}(a), each antenna element is attached to a flexible RF interface and can be repositioned within a confined region, typically a few wavelengths, without interrupting the RF chain \cite{zhu2025tutorial}. This repositioning is orchestrated by a central controller coordinating both motion actuation and baseband signal processing. Various hardware technologies, precision linear actuators, MEMS-based sliders, liquid-metal channels, or compliant robotic structures, provide continuous sub-wavelength positioning accuracy \cite{ning2025movable}. Controlled physical motion grants the transceiver several powerful capabilities: i) reshape its illumination and reception geometry, ii) navigate toward strong multipath clusters, iii) escape deep fading regions, and iv) adaptively enlarge its effective aperture for sensing.

MAS architectures differ in their DoFs \cite{li2025movable}.
Fig.~\ref{fig3}(b) illustrates representative designs: 1DMA enabling linear motion, 2DMA enabling planar scanning, 3DMA enabling volumetric beam alignment, and 6DMA additionally enabling rotational DoF. These architectures progressively enhance spatial controllability at the cost of increasing actuation complexity.  By dynamically adjusting both position and orientation, MAS establishes a tunable spatial interface between the transceiver and its surrounding electromagnetic environment, transforming antennas from passive endpoints into active participants in channel reconfiguration. In essence, MAS transforms channel propagation from a static and uncontrollable phenomenon into a tunable spatial resource, enabling unprecedented adaptability for dynamic wireless environments.

\begin{figure}[t]
    \centering
    \includegraphics[width=0.8\linewidth]{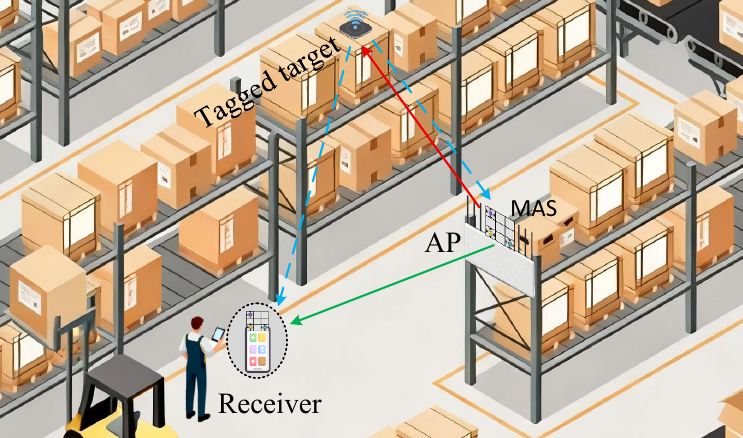}
    \caption{The MAS-empowered B-ISAC architecture.}
    \label{fig4}
\end{figure}

\subsection{Architecture of MAS-empowered B-ISAC}

As illustrated in Fig.~\ref{fig4}, an MAS-assisted B-ISAC system consists of three cooperative components that jointly operate over a geometry-aware two-hop link. The full-duplex MAS-enabled AP functions as a steerable illuminator, not only designing the ISAC waveform but also physically adjusting its aperture within a confined region to reshape the forward illumination toward the tagged target.  Unlike conventional fixed arrays that rely solely on electronic beam steering over a static aperture, this physical displacement shifts the effective phase center and alters the underlying propagation geometry. The tagged target, attached to objects of interest, remains an ultra-low-power tag, reflecting the incident ISAC waveform while embedding its modulated information, thereby acting simultaneously as a communication node and a geometry-aware sensing reflector. On the reception side, the MAS-enabled receiver (whether co-located with the AP or separate) behaves as an agile listener, repositioning its antennas to align with dominant return paths and avoid local fading dips, thereby maximizing the capture of information-rich echoes.

Together, these elements form forward illumination and backward sensing-communication paths whose geometry can be adaptively adjusted at the transceiver side as the channel evolves. This introduces controllable spatial degrees of freedom that allow the system to reshape the transmitter-tag-receiver coupling over time, alleviating geometric fragility and laying the foundation for the operating principles discussed next.

\subsection{Principles of MAS-empowered B-ISAC}

Integrating MAS into B-ISAC fundamentally transforms the cascaded channel from a static condition to be endured into a dynamic resource to be commanded. Instead of accepting geometry as a fixed deployment parameter, MAS elevates it to a tunable element of the system. The operating principles can be understood across four tightly coupled layers.

\subsubsection{Physical-Layer Principle}
Sub-wavelength antenna displacements perturb the cascaded channel by modifying phase accumulation and path-length differentials along both the forward and backward links. Because the backscatter channel is multiplicative, these minute perturbations produce amplified effects on the composite gain and phase. This motion-induced channel plasticity is the most fundamental mechanism through which physical movement becomes an effective RF control parameter.

\subsubsection{Geometric Principle}
Antenna motion directly reshapes the incidence and return angles at the tag, to which the backscatter process is highly sensitive. By adjusting positions within a wavelength-scale region, the system can realign the illumination direction with the tag’s reflective axis and recover favorable angle pairs that maximize cascade gain. This motion-angle coupling converts geometric alignment from a static deployment constraint into a controllable variable.

\subsubsection{Channel Principle}
Controlled motion empowers the transceiver to become a sculptor of the multipath environment. It can proactively probe the spatial domain, then reinforce desirable components (like a strong LoS or a valuable NLoS path) and suppress interfering ones. As the aperture moves, it guides the channel into a new, more favorable configuration. This process creates a powerful motion-enabled angular diversity, effectively helping the system hop out of deep fades and navigate around obstructions to maintain link stability in dynamic or cluttered settings.

\subsubsection{System Principle}
The physical, geometric, and channel-level effects are integrated and executed through a closed-loop spatial intelligence mechanism. The transceiver moves, observes the updated backscatter response, computes an optimal position, and moves again. This iterative perceive-and-optimize cycle establishes real-time geometric adaptability, replacing the spatial rigidity of conventional B-ISAC with a continuously tunable spatial interface that actively seeks out the best possible connection.

\subsection{Capabilities of MAS-empowered B-ISAC}

While the previous subsection outlined the physical and geometric principles that enable MAS, their impact is ultimately manifested at the system level. MAS elevates B-ISAC from a geometry-limited architecture to one that can actively rewrite its own propagation conditions. The resulting advantages are profound and can be summarized as follows.

\subsubsection{Geometry-aware Channel Reconfiguration}
Because the cascaded backscatter link is highly angle-sensitive, even small variations in tag pose or blockage can severely degrade performance. MAS mitigates this fragility by repositioning illumination and reception points to restore favorable geometric configurations. This enables the system to maintain strong LoS or high-quality NLoS components as the environment changes, significantly improving communication and sensing robustness.

\subsubsection{Motion-enhanced Sensing Fidelity}
Conventional B-ISAC inherits the fixed viewpoint of passive radar, which limits spatial resolution. MAS overcomes this limitation by capturing multiple micro-displaced observations, effectively enlarging the aperture and introducing angular diversity. These spatially enriched measurements sharpen range-angle estimation, improve multipath separability, and enhance robustness against clutter and occlusions, leveraging the geometric structure embedded in the backscattered waveform.

\subsubsection{Synergistic Integration of Motion and Backscatter}
MAS and passive backscatter modulation reinforce each other. Controlled transceiver movement strengthens tag-to-receiver coupling for communication, enhances the spatial signatures needed for tag localization or environmental inference, and aligns the optimal geometry for joint sensing and communication. Motion control thus becomes an additional optimization dimension, interacting with waveform design and signal processing to form a unified spatial-digital adaptation loop.

\begin{figure*}[t]
    \centering
    \vspace{-5mm}
    \includegraphics[width=0.68\textwidth]{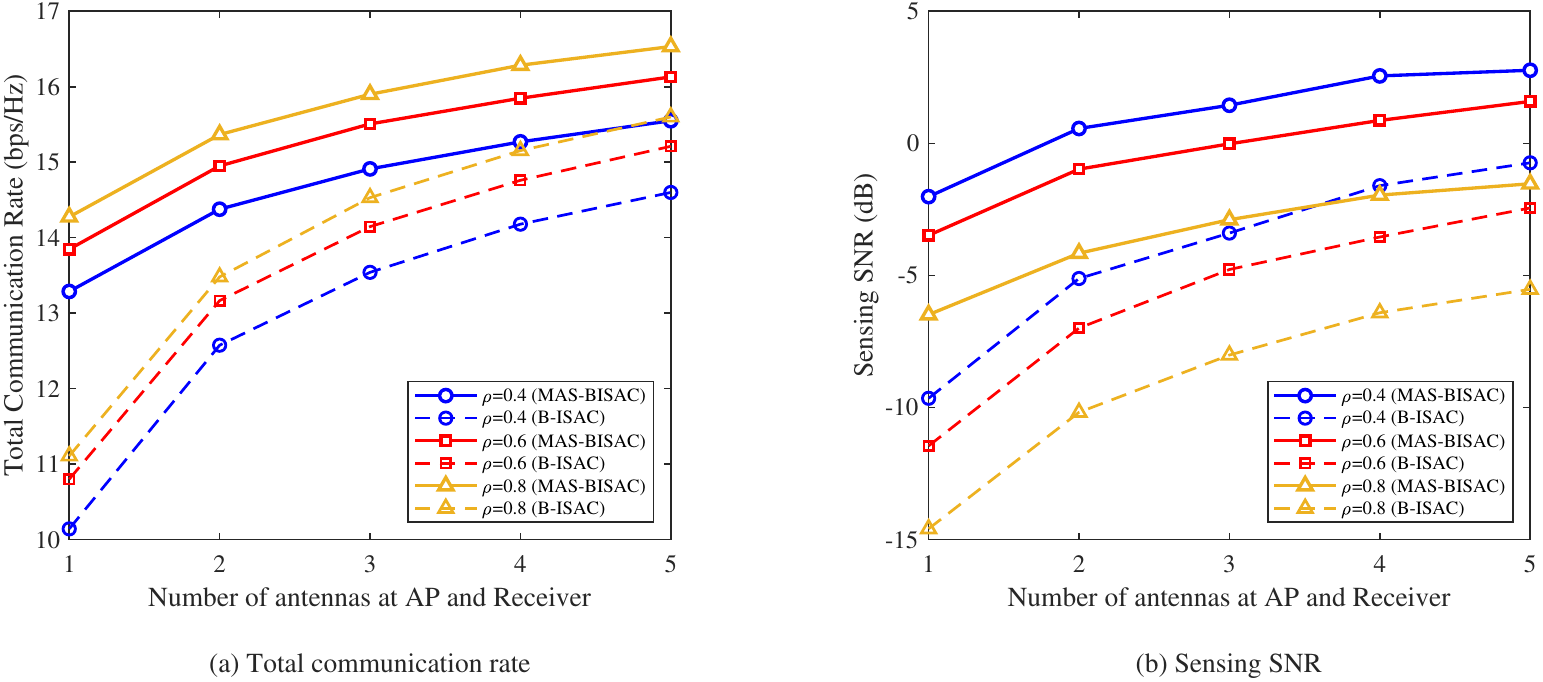}
        \caption{The communication-sensing performance under different the number of antennas and power-splitting factors.}
    \label{fig5}
\end{figure*}

\subsection{Numerical Results}
To illustrate the tangible gains brought by antenna mobility, we numerically compare the proposed MAS-assisted B-ISAC architecture with a conventional static B-ISAC baseline. The evaluation adopts a 3GPP urban microcell scenario with additive white Gaussian noise, where both systems operate under the same bandwidth and total transmit power while performing joint communication and sensing. For fairness, the only additional DoF in the MAS-assisted case is the ability of the AP/Receiver antennas to move within a small (sub-wavelength) region and adapt their positions based on the instantaneous channel state.

As shown in Fig.~\ref{fig5}(a) and Fig.~\ref{fig5}(b), introducing antenna motion consistently improves system performance across various antenna configurations and power-splitting factors (i.e.,  $\rho$). By adaptively repositioning antennas to avoid double-path fading and optimize the reflection geometry, the MAS-assisted B-ISAC achieves higher communication rates and stronger sensing SNRs than its static counterpart.  This confirms the efficiency of antenna mobility as a lever for enhancing integrated communication-sensing performance in B-ISAC systems.

\begin{figure*}[t]
\vspace{-5mm}
    \centering
    \includegraphics[width=0.8\textwidth]{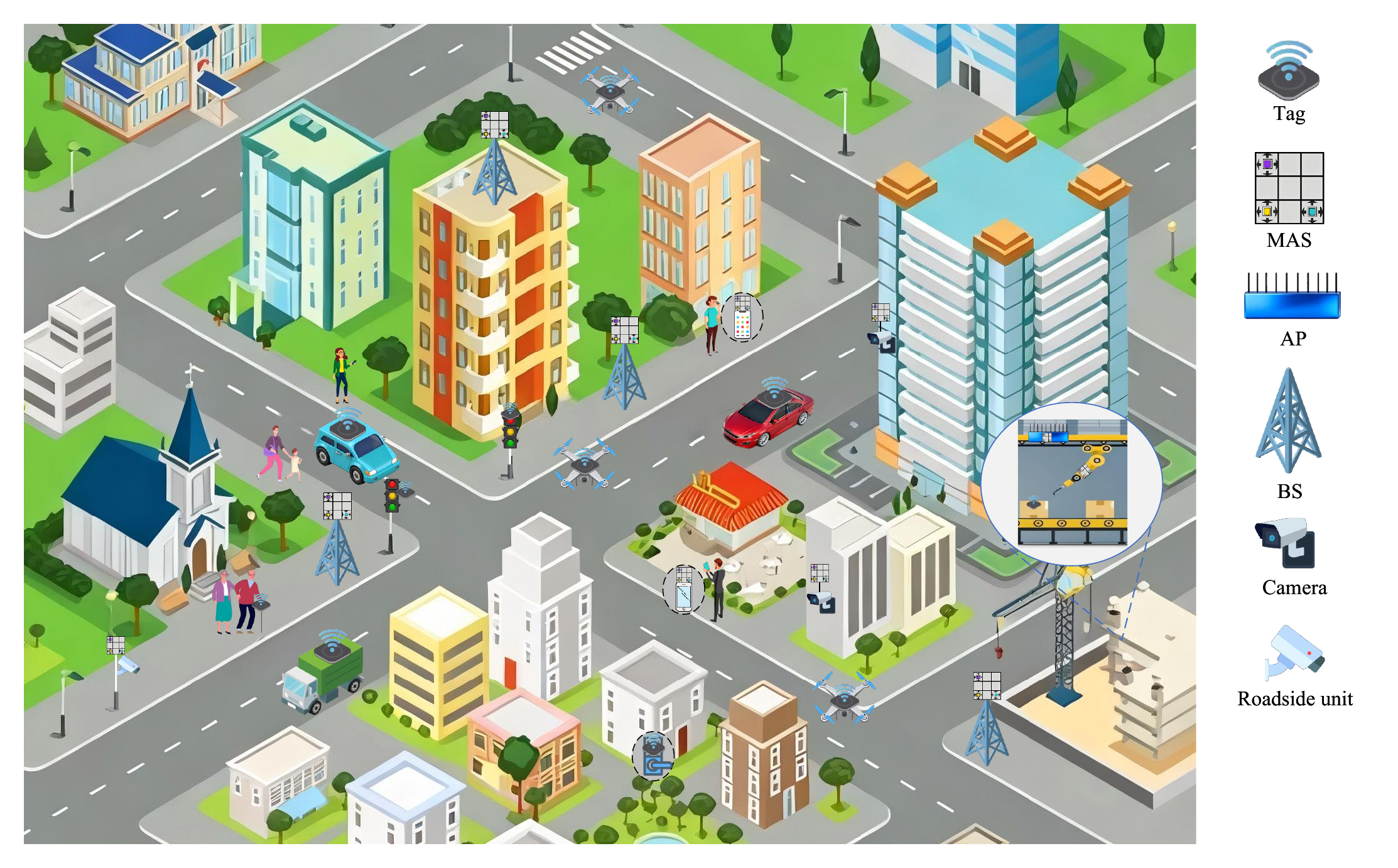}
    \caption{The applications of B-ISAC systems}
    \label{fig6}
\end{figure*}

\section{The Application of MAS-empowers B-ISAC}

The MAS-empowered B-ISAC architecture introduces transceiver-side spatial adaptability that directly addresses the geometric fragility of conventional B-ISAC. The following representative scenarios show where such adaptability becomes essential for stable and high-fidelity joint communication-sensing.

\subsection{Urban Transportation}
Modern transportation systems, such as autonomous vehicles, roadside perception units, and V2X infrastructures, depend on stable interaction with vehicle-mounted tags amid constantly shifting propagation conditions. Dense buildings, large vehicles, and sharp turning maneuvers routinely disrupt the cascaded link, leading to abrupt drops in communication reliability and sensing accuracy for fixed-geometry B-ISAC. MAS alleviates these disruptions by allowing roadside units to perform sub-wavelength adjustments that preserve favorable illumination and rapidly reacquire dominant return paths after transient blockage. Such micro-motions smooth Doppler fluctuations, stabilize lane-change dynamics, and maintain sensing fidelity across intersections and tunnels. In effect, MAS upgrades static roadside receivers into geometry-adaptive nodes capable of sustaining lane-level reliability throughout dense urban corridors.

\subsection{Industrial Logistics and Warehousing} 

Warehouses and logistics centers expose B-ISAC systems to severe multipath, metal-induced fading, and fast-changing object layouts. Static readers are often trapped in blind zones formed by racks, aisles, and conveyor systems, making it difficult to maintain continuous sensing or tag detection. MAS provides an effective spatial remedy. Small movements, such as sliding along a rack edge, lifting above a metal shelf, or shifting laterally near a conveyor path, help recover blocked LoS, avoid destructive nulls, and obtain more distinguishable multipath profiles. This spatial agility ensures uninterrupted inventory tracking, precise object localization, and robust coordination among mobile robots, forming a foundation for highly automated, self-optimizing logistics operations.

\subsection{Human-centric Sensing} 

Wearable and implantable backscatter devices enable continuous motion and health monitoring, yet human-centered environments generate highly time-varying scattering. Arm swings, fabric motion, posture shifts, or limb occlusions frequently place the tag in unfavorable geometric configurations, degrading both the received backscatter and the derived sensing features. MAS addresses this challenge by physically realigning its sensing aperture as users move. Even slight displacements can restore a strong path after transient shadowing, capture richer micro-motions such as breathing or gait patterns, and isolate gesture signatures from ambient clutter. These capabilities make MAS-assisted B-ISAC particularly well suited for continuous e-health monitoring, smart fitness analytics, and safety-aware human-robot interaction.

\subsection{Low-altitude Economy and Aerial IoT}

The emerging low-altitude economy, covering drone logistics, aerial inspection, and air-ground IoT, involves rapidly moving tagged platforms in highly dynamic 3D channels. Fast rotations, altitude changes, and intermittent blockage from buildings or terrain cause sharp variations in the forward and backward backscatter paths, making static B-ISAC unreliable for aerial tracking and sensing. MAS adds the needed spatial agility: ground stations can slightly adjust antenna positions to maintain favorable illumination and quickly recover strong returns after occlusions, while MAS-equipped aerial nodes can optimize their apertures relative to ground tags or infrastructure. These controlled micro-motions stabilize the cascaded channel, improve altitude and velocity estimation, and support robust air-ground interaction, positioning MAS-assisted B-ISAC as a key player for drone traffic management and aerial IoT services.

\section{Future Directions toward Practical MAS-empowered B-ISAC}

Realizing deployable MAS-assisted B-ISAC systems requires moving beyond idealized geometric models toward robust designs that jointly consider motion, backscatter, and sensing. The following directions outline the key challenges and opportunities that must be addressed for practical, scalable deployment across real environments.

\subsection{Channel Modeling and Environment Reconstruction}

In MAS-assisted B-ISAC, the cascaded backscattering channel is no longer determined solely by static geometry but also by antenna trajectories and sub-wavelength displacements. This calls for motion-aware channel models that capture how small movements reshape forward and backward links, load-modulated reflections, and time-varying multipath. Embedding these models into electromagnetic digital twins is a promising direction, enabling the prediction of how motion impacts fading and sensing resolution. The main challenge is achieving sufficient fidelity without sacrificing real-time computability in dynamic urban, industrial, or human-centric scenes.

\subsection{Joint Motion Control and Resource Management }
Once motion becomes a controllable DoF, the design focus shifts from optimizing waveforms for a fixed channel to jointly optimizing waveforms and motion so that the channel itself becomes favorable. Future frameworks need to co-design antenna trajectories, beamforming, power splitting, and tag load modulation under mechanical constraints, actuation delay, and channel coherence limits. Control-theoretic motion planning and learning-based strategies, such as reinforcement learning, are natural candidates, but must carefully account for motion energy and latency so that the system can decide when to move, when to beamform, and when to rely purely on baseband processing.

\subsection{Scalable Multi-tag and Multi-MAS Coordination}

Real deployments will feature large numbers of tags and potentially multiple MAS-enabled BSs or APs with overlapping coverage. Antenna trajectories determine which tags can be illuminated reliably and which echoes remain separable, while tag activity and traffic patterns influence where antennas should move next. This tight coupling between physical motion and MAC-layer scheduling demands scalable coordination schemes that avoid excessive control signaling and synchronization overhead. A key goal is to treat spatial adaptability as a network-wide resource, not a local feature confined to a single transceiver.

\subsection{Reliability and Compatibility}

Continuous antenna motion alters channel statistics and, if unmanaged, can disrupt both communication reliability and sensing accuracy. Practical MAS-assisted B-ISAC systems must incorporate motion-aware channel prediction, robust waveform and receiver designs tolerant to spatial perturbations, and closed-loop control policies that prevent abrupt performance drops during movement. At the same time, they must remain compatible with emerging architectures such as IRS-assisted links, UAV-mounted nodes, and cell-free networks, relying on unified control interfaces and coexistence mechanisms that allow MAS to be introduced without redesigning existing ISAC infrastructures or legacy passive tags.

\subsection{Security and Privacy}

Controllable spatial motion creates new opportunities and risks for physical-layer security. On the positive side, motion-induced geometric diversity can be exploited for artificial noise generation, key extraction from rapidly varying channels, and agile defenses against spoofing or jamming. On the negative side, motion patterns and backscattered signals may leak information about user locations, operational routines, or personal activities, and motion-control channels themselves may become attack targets. Future work must therefore develop motion-aware security primitives and privacy-preserving sensing schemes that jointly consider RF behavior and mechanical dynamics, especially in human-centric deployments.

\section{Conclusion}

This article has presented a MAS-assisted B-ISAC paradigm as a decisive strategy to overcome the geometric bottlenecks of conventional passive systems. By turning the transceiver geometry into a controllable variable, MAS-empowered B-ISAC effectively mitigates double fading and enhances robustness, thereby improving communication reliability and sensing fidelity while strictly preserving the ultra-low-power and spectral efficiency advantages of BackCom. Numerical results and scenario-based analyses provide proof-of-concept evidence, showing substantial gains in challenging urban, industrial, low-altitude, and human-centric environments. Looking ahead, practical deployment calls for deeper research on motion-aware channel modeling, joint motion control and resource management, scalable coordination mechanisms, and reliability and security. With progress on these fronts, MAS-assisted B-ISAC is well positioned to evolve from a conceptual paradigm into a key architectural component of adaptive, low-power, sensing-native 6G networks.

\end{document}